\documentclass[useAMS,usenatbib]{mn2e}

\usepackage[psamsfonts]{amssymb}
\usepackage[dvips]{graphicx}
\usepackage{amsmath,alltt}
\usepackage{multirow}
\usepackage{rotating}
\usepackage{lscape}

\title[Dynamical Evolution of Star Clusters]{The Effect of Orbital Eccentricity on the Dynamical Evolution of Star Clusters}
\author[Webb, J. J.]{Jeremy J. Webb$^{1}$, Nathan Leigh$^{2}$, Alison Sills$^{1}$, William E. Harris$^{1}$, Jarrod R. Hurley $^{3}$
\thanks{E-mail: webbjj@mcmaster.ca (JW), nleigh@rssd.esa.int (NL)} \\
$^{1}$McMaster University, Department of Physics and Astronomy, 1280 Main St. W., Hamilton, Ontario, Canada, L8S 4M1 \\
$^{2}$European Space Agency, Space Science Department, Keplerlaan 1,
2200 AG Noordwijk, The Netherlands \\
$^3$Centre for Astrophysics and Supercomputing, Swinburne University of Technology, P.O. Box 218, VIC 3122, Australia}

\begin{document}

\pagerange{\pageref{firstpage}--\pageref{lastpage}} \pubyear{2013}

\maketitle

\label{firstpage}

\begin{abstract}

We use $N$-body simulations to explore the influence of orbital eccentricity on the dynamical evolution of star clusters. Specifically we compare the mass loss rate, velocity dispersion, relaxation time, and the mass function of star clusters on circular and eccentric orbits.  For a given perigalactic distance, increasing orbital eccentricity slows the dynamical evolution of a cluster due to a weaker mean tidal field. However, we find that perigalactic passes and tidal heating due to an eccentric orbit can partially compensate for the decreased mean tidal field by energizing stars to higher velocities and stripping additional stars from the cluster, accelerating the relaxation process. We find that the corresponding circular orbit which best describes the evolution of a cluster on an eccentric orbit is much less than its semi-major axis or time averaged galactocentric distance. Since clusters spend the majority of their lifetimes near apogalacticon, the properties of clusters which appear very dynamically evolved for a given galactocentric distance can be explained by an eccentric orbit. Additionally we find that the evolution of the slope of the mass function within the core radius is  roughly orbit-independent, so it could place additional constraints on the initial mass and initial size of globular clusters with solved orbits. We use our results to demonstrate how the orbit of Milky Way globular clusters can be constrained given standard observable parameters like galactocentric distance and the slope of the mass function. We then place constraints on the unsolved orbits of NGC 1261,NGC 6352, NGC 6496, and NGC 6304 based on their positions and mass functions.

\end{abstract}


\section{Introduction} \label{intro}

Massive star clusters in the Milky Way (MW), called globular clusters (GCs), have typical total masses and ages ranging from $\sim$ 10$^4$ - 10$^6$ M$_{\odot}$ and $\sim$ 10-12 Gyrs, respectively \citep{harris96, marinfranch09}. They have had time for their structural properties and stellar mass functions (MFs) to have been modified from their primordial forms due to both stellar evolution and stellar dynamics. Thus, in order to constrain the initial cluster conditions and mass function, simulations are needed to rewind their dynamical clocks.

The dominant mechanisms which drive the dynamical evolution of star clusters are:

\begin{itemize}

\item Stellar Evolution
\item Two-body Relaxation
\item Tidal Stripping
\item Tidal Heating
\item Disk Shocking

\end{itemize}

Stellar evolution is initially the main driver of dynamical evolution in a cluster as significant mass loss occurs when massive stars quickly evolve off the main sequence and go supernova.  After 2-3 Gyr, two-body relaxation, the cumulative effects of long-range gravitational interactions between stars acting to alter stellar orbits within the cluster, becomes dominant \citep[e.g.][]{henon61, henon73, spitzer87, heggie03, gieles11}. The most massive stars accumulate in the central cluster regions, and the lowest mass stars are dispersed to wider orbits.  The re-distribution of low and high mass stars, known as mass segregation, is also a source of mass loss with the probability of ejection past the tidal boundary increasing with decreasing stellar mass. Therefore, two-body relaxation will slowly modify the distribution of stellar masses within clusters, and can cause very dynamically evolved clusters to appear severely depleted of their low-mass stars \citep[e.g.][]{vonhippel98, koch04, demarchi10}.

Tidal stripping is the removal of stars from a cluster by the host galaxy. The galactic potential imposes a theoretical boundary around a globular cluster, known as the tidal radius $r_t$ or the Jacobi radius $r_J$. Beyond $r_t$, a star will feel a greater acceleration towards the galaxy center than it feels towards the center of the cluster, and will therefore escape \citep{binney08}. For clusters subject to a strong tidal field, stripping serves to both accelerate mass loss and minimize cluster size.

Tidal heating is an effect only experienced by clusters which experience a non-static tidal field, and so only applies to clusters with eccentric orbits or circular orbits in non-spherically symmetric potentials. The non-static tidal field injects energy into the stellar population of a globular cluster and the kinetic energy of individual stars increases. Energy injection leads to both the energization of stars to larger orbits and the ejection of stars that would otherwise remain bound to the cluster. The effects of energy injection are strongest during a perigalactic pass where the cluster experiences a sudden and dramatic increase in the local potential \citep{spitzer87, webb13}. Disk shocking is a specific and extreme form of tidal heating, similar to a perigalactic pass, as the local potential changes dramatically when the cluster passes through the Galactic disk.

While stellar evolution, two-body relaxation and tidal stripping have all been well studied for GCs in isolation and on circular orbits in realistic potentials, how these mechanisms change as a function of orbital eccentricity remains unclear. The purpose of this study is to determine how tidal heating, due to a non-circular orbit in a disk potential, and energy injection during perigalactic passes can influence both relaxation and mass loss due to tidal stripping. All of the Galactic GCs with solved orbits are non-circular \citep{dinescu99, dinescu07, dinescu13}, therefore understanding the effects of orbital eccentricity are key to any future studies of GCs.

We evolve model $N$-body clusters for 12 Gyr with a range of orbits in a Milky Way-like potential. Clusters with different orbits experience different degrees of tidal stripping and tidal heating, which can have significant effects on both the low-mass stellar population in the outer regions of the cluster and cluster density. In Section~\ref{nbody} we discuss the $N$-body models used in this paper. To study how orbital eccentricity can alter the dynamical evolution of a cluster, we investigate the effect that tidal heating has on cluster mass loss rate (Section~\ref{mloss}), velocity dispersion (Section~\ref{vdis}), relaxation time (Section~\ref{strh}), and the stellar MF (Section~\ref{smfunc}). Within Section~\ref{smfunc}, the evolution of the MF in different regions of the cluster is also discussed. Finally in Section~\ref{mwapp}, we illustrate how present day characteristics of GCs can be used to provide constraints on cluster orbits. We then place constraints on the orbits of specific GCs that remain unsolved. We summarize our results in Section~\ref{summary}.  

\section{N-body models} \label{nbody}

We use the NBODY6 direct $N$-body code \citep{aarseth03} to study the evolution of model star clusters over 12 Gyr. The models in this study begin with 96000 single stars and 4000 binaries and have a total initial mass of $6 \times 10^4 M_\odot$. Since we are only concerned with the influence of orbital eccentricity on cluster evolution, only the initial position and initial velocity vary from model to model while all other parameters remain unchanged. 

A \citet{kroupa93} IMF between 0.1 and 30 $M_{\odot}$ is used to assign masses to individual stars, all with a metallicity of $Z=0.001$. For binary stars, the total mass of the binary is set equal to the mass of two randomly selected stars. The mass-ratio between the primary and secondary masses is then randomly selected from a uniform distribution. The distribution of \citet{duquennoy91} is used to set the initial period of each binary and orbital eccentricities are assumed to follow a thermal distribution \citep{heggie75}. Initial positions and velocities of the stars are based on a Plummer density profile \citep{plummer11,aarseth74} with a cut-off at $\sim 10 \ r_m$ to avoid the rare case of stars positioned at large cluster-centric distances. The initial half-mass radius $r_{m,i}$ of each model is 6 pc. The algorithms for stellar and binary evolution are described in \citet{hurley08a, hurley08b}.

The Galactic potential is made up of a $1.5 \times 10^{10} M_{\odot}$ point-mass bulge, a $5 \times 10^{10} M_{\odot}$ \citet{miyamoto75} disk (with $a=4.5\,$kpc and $b=0.5\,$kpc), and a logarithmic halo potential \citep{xue08}. The combined mass profiles of all three components force a circular velocity of 220 km/s at a galactocentric distance of $8.5\,$kpc. The incorporation of the Galactic potential into NBODY6 is described by \citet{aarseth03} and \citet{praagman10}. In order for the model clusters to experience a spherically symmetric tidal field they were set to orbit in the plane of the disk, eliminating factors such as disk shocking or tidal heating due to a non-spherically symmetric potential.

Since we are only focussed on stars that are energetically bound to the cluster, the simulation eliminates stars with $r > 2 \ r_t$, where $r_t$ is the \citet{king62} tidal radius. We then calculate the total energy of each star given its kinetic energy, the potential energy due to all other stars in the cluster, and the tidal potential \citep{bertin08, webb13}. Stars with $E > 0$ are considered to be unbound, and are not included in calculations of cluster parameters. It should be noted that a star with $E > 0$ can be recaptured at a later time if it does not travel beyond $2 \ r_t$.

We first simulate three clusters with orbital eccentricities of 0 (circular orbit), 0.5, and 0.9, where eccentricity is defined as $e = \frac{R_{a}-R_p}{R_{a}+R_p}$. $R_{a}$ and $R_{p}$ are the apogalactic and perigalactic distance of the orbit, respectively. All three models have an $R_{p}$ equal to 6 kpc and are located at $R_p$ at time zero. For comparison purposes we also simulate two additional models with circular orbits at the apogalacticon of the $e = 0.5$ and $e = 0.9$ models, corresponding to orbits at 18 kpc and 104 kpc, respectively. Therefore we can directly compare the properties of a cluster on an eccentric orbit to clusters on circular orbits at both $R_p$ and $R_a$. 

The initial model parameters are summarized in Table \ref{table:modparam}, with model names based on orbital eccentricity (e.g. e05) and either circular radius or radius at apogalacticon (e.g. r18).

\begin{table}
  \caption{Model Input Parameters}
  \label{table:modparam}
  \begin{center}
    \begin{tabular}{lcccc}
      \hline\hline
      {Model Name} & {$r_{m,i}$} & {$R_p$} & {$v_p$} & {e} \\
      { } & {pc} & {kpc} & {km/s} { } \\
      \hline

e0r6 & 6 & 6 & 212 & 0 \\
e05r18 & 6 & 6 & 351.5 & 0.5 \\
e0r18 & 6 & 18 & 232 & 0 \\
e09r104 & 6 & 6 & 543.5 & 0.9 \\
e0r104 & 6 & 104 & 225.25 & 0 \\

      \hline\hline
    \end{tabular}
  \end{center}
\end{table}

\section{Mass Loss Rate} \label{mloss}

\begin{figure}
\centering
\includegraphics[width=\columnwidth]{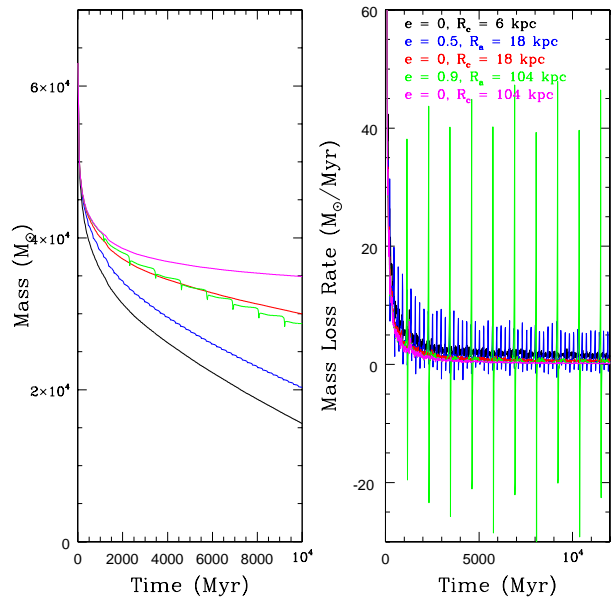}
\caption{Mass (left) and mass loss rate (right) of each model cluster as a function of time. Models are separated by colour as indicated.}
  \label{fig:mloss}
\end{figure}

The most important characteristic of a globular cluster is its total mass, as it sets $r_t$, the relaxation time $t_{rh}$ and velocity dispersion $\sigma_V$ of the cluster. Since our models all start with the same initial mass, the key feature which sets the models apart is their mass loss rate. Mass loss due to stellar evolution will be identical from model to model, however mass loss due to tidal stripping is orbit dependent since $r_t$ is a function of the instantaneous galactocentric distance $R_{gc}$ of a cluster. The total mass (left panel) and mass loss rate (right panel) of each model is plotted in Figure \ref{fig:mloss}. 

In Figure \ref{fig:mloss}, the mass loss rate of a GC on a circular orbit increases with decreasing $R_{gc}$, resulting in the present day mass of inner clusters (e0r6) to be much less than outer clusters (e0r104). The relationship between mass loss rate and $R_{gc}$ is expected as $r_t$ decreases linearly with $R_{gc}$. A stronger tidal field and smaller $r_t$ results in outer stars being easily removed from the cluster. The only exception to this rule is when a cluster is not tidally filling. 

\begin{figure}
\centering
\includegraphics[width=\columnwidth]{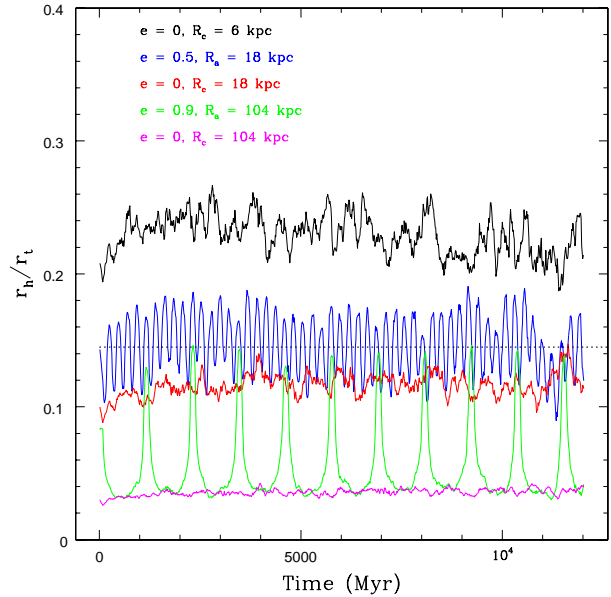}
\caption{Ratio of $\frac{r_h}{r_t}$ as a function of time. Models are separated by colour as indicated. The dotted line indicates a value of 0.145.}
  \label{fig:rfill}
\end{figure}

As shown in \cite{webb13}, clusters fill their instantaneous tidal radius at all times, independent of their orbital phase. That is to say there will always be energetically bound stars at or near $r_t$. However the degree to which a cluster is tidally filling depends on the ratio $\frac{r_h}{r_t}$, where a cluster can be approximated to be tidally filling if $\frac{r_h}{r_t} > 0.145$ \citep{henon61}. The fraction $\frac{r_h}{r_t}$ indicates whether the bulk of the cluster is centrally concentrated and only a few outer stars are affected by the tidal field (tidally under-filling) or if stars are more uniformly spread out between the cluster center and $r_t$. $\frac{r_h}{r_t}$ is plotted as a function of time for each model cluster in Figure \ref{fig:rfill}.

Tidally under-filling clusters, like e0r014, will therefore have a lower mass loss rate at a given $R_{gc}$ than if $\frac{r_h}{r_t} > 0.145$. Mass loss in under-filling GCs is primarily driven by stellar evolution and close two-body interactions occurring primarily in the dense cluster core.

The mass loss rate of a GC on an eccentric orbit can be much higher than if it had a circular orbit where it is currently observed, which is most likely near $R_a$. For example, in the left panel of Figure \ref{fig:mloss} the final masses of e05r18 and e09r104 are significantly less than the apogalactic cases of e0r18 and e0r104 respectively. So despite spending the majority of its lifetime near $R_a$, an eccentric cluster will be lower in mass than a cluster with a circular orbit at $R_a$. Periodic episodes of enhanced mass loss (right panel of Figure \ref{fig:mloss}) during a perigalactic pass are greater than the mass gained from recapturing stars as the instantaneous $r_t$ increases while the GC travels to $R_a$. 

It is interesting to note that e09r104 has a lower mass loss rate than e0r18 during the majority of its orbit, but e09r104 undergoes periodic episodes of mass loss at $R_p$ that results in similar mass profiles during the first 12 Gyr of their lifetime. e0r18 and e09r104 having similar mass profiles is in disagreement with the relationship between dissolution time and cluster orbit given by \citet{baumgardt03}. The results of \citet{baumgardt03} suggest that a cluster with an orbital eccentricity of 0.9 and perigalactic distance of 6 kpc would behave as if it had a circular orbit between 10.5 and 11.5 kpc and that e0r18 will take between 1.4 and 1.7 times longer to reach dissolution than e09r104. However, evolving our model clusters beyond 12 Gyr and defining the dissolution time as the time it takes for clusters to reach $35\%$ of their initial mass, we find that the mass profiles eventually diverge and e0r18 takes 1.35 times longer to reach dissolution than e09r104. The slight discrepancy between our models and the results of \citet{baumgardt03} can easily be attributed to our clusters having different initial conditions and orbiting in a different tidal field than those presented in \citet{baumgardt03}. e09r104 having a similar mass profile to e0r18 can be attributed to the clusters undergoing non-linear mass loss rates which result in both models losing similar amounts of mass over the first 12 Gyr of cluster evolution and different amounts of mass beyond 12 Gyr. Therefore we consider e09r104 to have an \textit{effective circular orbit $R_e$} near 18 kpc. $R_e$ can be thought of qualitatively as the circular orbit distance that an eccentric cluster could have and undergo the same dynamical evolution. \footnote{Unfortunately, no quantitative relationship between the orbit of e09r104 and its apparent $R_e$ of 18 kpc could be established.} 

e09r104 has a semi-major axis of 60 kpc and a time average galactocentric distance (\textit{$<R_{gc}>=\frac{1}{12 Gyr} \int_0^{12 Gyr} R_{gc}(t) dt$}) of 73 kpc, both significantly larger than $R_e$. Even the time averaged galactic potential experienced by e09r104 (\textit{$<\Psi>=\frac{1}{12 Gyr} \int_0^{12 Gyr} \Psi(t) dt$}), which is the exact same as a cluster with a circular orbit at 62 kpc, is larger than $R_e$. The circular orbit distance which experiences the same $<\Psi>$ as an eccentric cluster will be referred to as \textit{$R_\Psi$}, such that $\Psi(R_\Psi) = <\Psi>$. Hence perigalactic mass loss leads to the mass loss rate of an eccentric cluster being higher than if the cluster had a circular orbit at $<R_{gc}> $, $R_\Psi$ or with the same semi-major axis.

It should be noted that we consider e0r6 and e05r18 to be tidally filling, while e09r104 is only tidally filling near $R_p$. e0r18 is marginally filling, so while it is still subject to the effects of the tidal field, tidal heating and stripping will be less efficient than in tidally filling clusters. e0r104 is the only cluster that can be considered to be truly tidally under-filling over 12 Gyr, and its evolution independent of the tidal field.

\section{Velocity Dispersion} \label{vdis}

\begin{figure}
\centering
\includegraphics[width=\columnwidth]{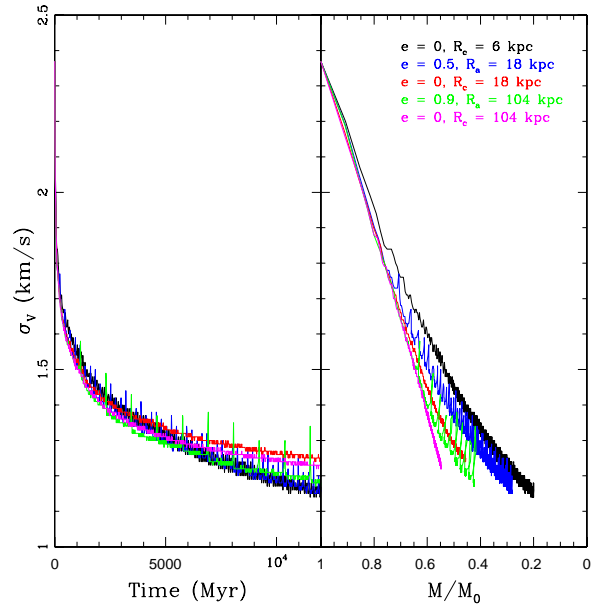}
\caption{Velocity dispersion as a function of time (left panel) and fraction of initial mass (right panel). Models are separated by colour as indicated.}
  \label{fig:sigv}
\end{figure}

An observable parameter that is commonly used to study the dynamical state of a globular cluster is its global line of sight velocity dispersion $\sigma_V$ (Equation \ref{sigv})

\begin{equation} \label{sigv}
\sigma_V=\sqrt{\frac{\sum\limits_{i=1}^N v_i^2}{N}}
\end{equation}

\noindent where $v_i$ is the line of sight velocity of individual stars. We have plotted the evolution of the global line of sight velocity dispersion of each model as a function of both time (left panel) and fraction of initial mass $\frac{M}{M_0}$ (right panel) in Figure \ref{fig:sigv}. The velocity dispersion was calculated along a random line of sight at each time step. Comparing model clusters as a function of fraction of initial mass is equivalent to comparing clusters on the same evolutionary timescale, as the fraction of initial mass lost from the system per relaxation time due to energy equipartition-driven dynamical evolution should be approximately the same for all clusters independent of their mass, as shown by \citet{lamers13}. It should be noted that since the model clusters are only simulated to 12 Gyr and not to dissolution, each model cluster will have lost a different fraction of its initial mass by the end of the simulation.

The trend is for the velocity dispersion of all models to decrease as they evolve, primarily due to mass loss over time. Since velocity dispersion is proportional to cluster mass and inversely proportional to size, both of which are dependent on orbit, it is difficult to relate velocity dispersion to cluster orbit when plotted as a function of time (Figure \ref{fig:sigv} left panel). However, if we plot velocity dispersion versus the fraction of initial mass (Figure \ref{fig:sigv} right panel) we are comparing clusters at the same mass. Since GC $r_h$ decreases with decreasing $R_{gc}$, we expectedly see a higher $\sigma_V$ for clusters with circular orbits that experience a stronger tidal field for a given fraction of initial mass.

While stronger Galactic tides increase the velocity dispersion of a GC on a circular orbit, tidal heating due to a non-circular orbit can play a secondary role. In Figure \ref{fig:sigv} we see that the velocity dispersion of GCs with eccentric orbits spikes during perigalactic passes as tidal heating injects all stars with additional energy \citep{spitzer87, gnedin99}, with the line of sight velocity dispersion deviating by up to 0.15 km/s and the three dimensional velocity dispersion deviating by up to 0.3 km/s. When this energy is injected into the cluster, the acceleration (and hence energy) imparted to these stars will push them outwards as they move closer to being energetically unbound and can even strip outer low-mass stars from the cluster if their initial binding energy is low enough, in agreement with \citet{webb13}. 

Even though the majority of the high velocity stars will escape the cluster and not be recaptured, some stars will remain bound. The periodic process of increasing the velocity dispersion during a perigalactic pass acts to slow the decrease in $\sigma_V$ compared to if it had a circular orbit at $<R_{gc}>$, $R_\Psi$, the semi-major axis of the eccentric cluster, or $R_a$. Therefore for two given clusters that are equal in mass at the same $R_{gc}$, a higher velocity dispersion will indicate an eccentric orbit assuming the eccentric cluster is located near apogalacticon.

\section{Relaxation} \label{strh}

We next wish to examine how cluster orbit affects the timescale over which the distribution of stellar energies approaches equilibrium, known as the relaxation time \textit{$t_{rh}$}  \citep{heggie03, trenti13}. $t_{rh}$ is given by Equation \ref{trh} \citep{meylan01}, where M is the total GC mass, $\bar{m}$ is the mean stellar mass, and $r_h$ is the half-light radius.

\begin{equation}\label{trh}
t_{rh}[yr]=(8.92 \times 10^5) \frac{(M / M_{\odot})^{\frac{1}{2}}}{(\bar{m} / M_{\odot})} \frac{(r_h / 1 pc)^{\frac{3}{2}}}{log(0.4 M / \bar{m})}
\end{equation}

\begin{figure}
\centering
\includegraphics[width=\columnwidth]{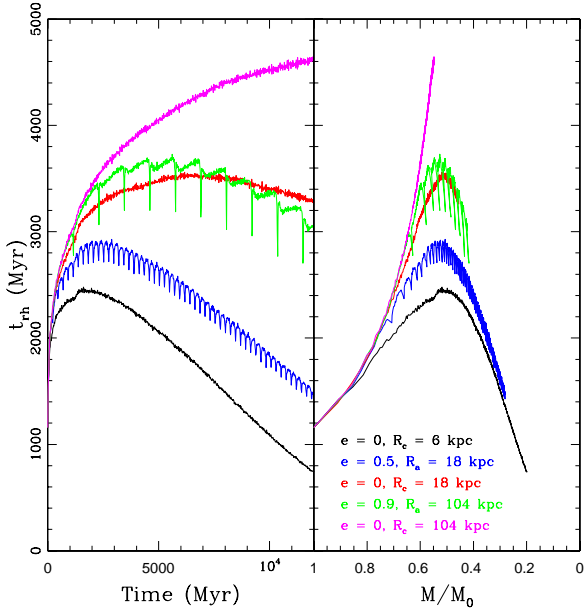}
\caption{Half-mass relaxation time of each model cluster as a function of time (left panel) and fraction of initial mass (right panel). Models are separated by colour as indicated.}
  \label{fig:trh}
\end{figure}

The relaxation time, plotted as a function of time (left panel) and fraction of initial mass (right panel) in Figure \ref{fig:trh}, is dependent on all three of the previously discussed cluster characteristics; mass, $r_h$, and velocity dispersion.

As previously discussed, a cluster which experiences a strong tidal field will have a higher mass loss rate, higher velocity dispersion and be smaller in size than a cluster which experiences a weaker tidal field. While a larger velocity dispersion will increase the relaxation time of a GC, differences in $\sigma_V$ due to cluster orbit are minimal compared to the differences in mass and size of clusters in different tidal fields. Therefore the relaxation and segregation times of a cluster are primarily dependent on cluster size and density, both of which are proportional to $R_{gc}$. With the exception of e0r104, $t_{rh}$ decreases with time after its initial expansion while each cluster loses mass and contracts. Since e0r104 is undergoing a near-zero mass loss rate and still expanding, $t_{rh}$ continues to increase.

Figure \ref{fig:trh} indicates that a cluster with an eccentric orbit relaxes on a timescale between that of GCs with circular orbits at $R_p$ and $R_a$. Increasing eccentricity increases $t_{rh}$ relative to the $R_p$ case, primarily due to the eccentric cluster having a larger $r_h$. Therefore for two clusters at the same $R_{gc}$, the cluster with a more eccentric orbit which brings it deeper into the galactic potential will have a shorter relaxation time and be more mass segregated than a cluster with a near-circular orbit. Similar to the evolution of total mass and $\sigma_V$ in Figures \ref{fig:mloss} and \ref{fig:sigv}, model e09r104 has a relaxation time profile that overlaps with e0r18.

\section{Evolution of the Mass Function} \label{smfunc}

The overall effect of orbital eccentricity on the dynamical evolution of GCs is observed in the stellar MF. Increased tidal stripping results in eccentric clusters being severely depleted of mass segregated low-mass stars compared to clusters with circular orbits near the same $R_{gc}$. Hence studying the stellar MF of a GC allows for constraints to be placed on its orbital eccentricity.

\subsection{Evolution of $\alpha$} \label{alpha}

We quantify the evolution of the MF by calculating the exponent $\alpha$, where $\alpha$ is defined in Equation \ref{dndm}. 

\begin{equation} \label{dndm}
\frac{dN}{dm} \propto m^\alpha
\end{equation}

\begin{figure}
\centering
\includegraphics[width=\columnwidth]{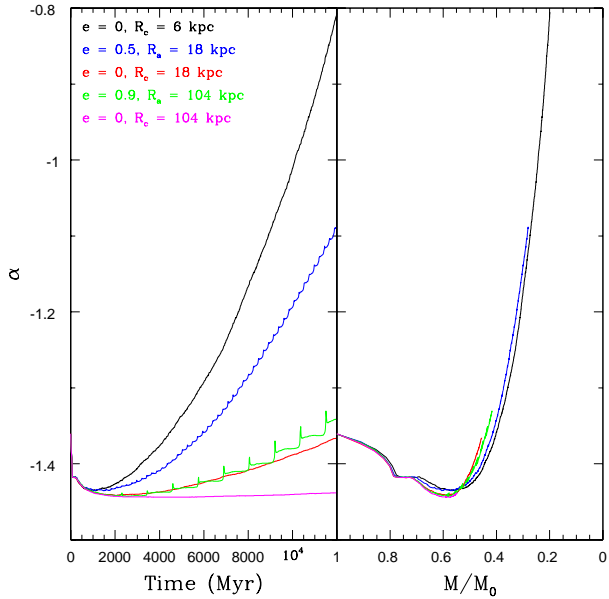}
\caption{ The evolution of the global $\alpha$ is plotted as a function of time (left panel) and fraction of initial mass (right panel). Models are separated by colour as indicated.}
  \label{fig:alpha}
\end{figure}

\noindent In this form, the traditional Salpeter initial MF has $\alpha$ = -2.35 \citep{salpeter55}. For each model, $\alpha$ is the best fit slope to a plot of $log(\frac{dN}{dm})$ versus $log(m)$, calculated over mass bins greater than $0.15 M_{\odot}$ and less than the main sequence turn-off. The evolution of the global $\alpha$ for each of our models is plotted in Figure \ref{fig:alpha} as a function of time (left panel) and fraction of initial mass (right panel).

Almost immediately, $\alpha$ decreases from its initial value due to both stellar evolution and the breaking up of binaries which are assumed to be unresolved. After 1000-2000 Myr $\alpha$ begins to increase as a function of time at a faster rate for GCs which experience a stronger tidal field. The accelerated evolution of $\alpha$ is a direct result of increased mass loss due to tidal stripping producing a lower mass cluster with a shorter relaxation time and a smaller scale size ($r_t \propto M^{\frac{1}{3}} R_{gc}^{\frac{2}{3}}$). As a function of fraction of initial mass, all models again undergo a similar initial evolution in $\alpha$. It is not until after the first 1000 Myr and each cluster has completed multiple orbits and experienced the combined effects of the galactic potential that the evolution of $\alpha$ becomes orbit dependent. For a given fraction of initial mass, $\alpha$ will then be higher for a cluster with a large $<R_{gc}>$ as the weaker tidal field can only remove the least massive of the low mass stars. A stronger tidal field can remove stars over a larger mass range, slowing the evolution of $\alpha$.

We have already shown that tidal heating, on top of the lower mass and smaller scale size of an eccentric cluster, accelerates its dynamical evolution compared to a GC with a circular orbit and either the same semi-major axis, the same $<R_{gc}>$ or the same $R_\Psi$. Comparing GCs as a function of initial mass, $\alpha$ increases at a faster rate with increasing eccentricity (for a given $R_p$) because the weaker tidal field again can only remove the lowest of low mass stars. Since clusters with higher orbital eccentricities are subject to increased tidal heating and a tidal shock at $R_p$, a larger fraction of low-mass stars populating the outer regions have the potential to be tidally stripped.

\subsection{Radial Dependence of the Mass Function}

It is often the case that the slope of the mass function for a given GC is measured in a specific region of the GC \citep[e.g.]{demarchi10}. Therefore, to properly compare with observable parameters we consider the evolution of $\alpha$ for stars in different radial regions of the cluster. Specifically we focus on stars within the $10 \%$ Lagrangian radius ($r_{10}$), stars between $r_{10}$ and the half mass radius ($r_m$), and bound stars beyond $r_m$. For our purposes, $r_m$ is used as a substitute for $r_h$ because it undergoes a smoother evolution from time step to time step than $r_h$.

\begin{figure}
\centering
\includegraphics[width=\columnwidth]{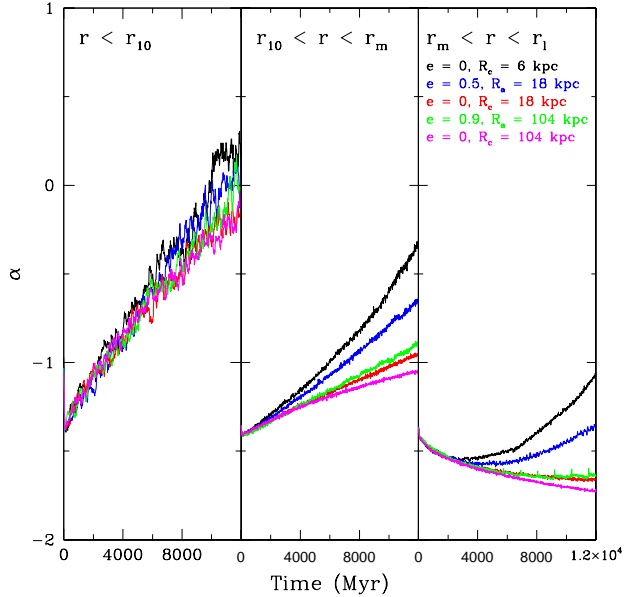}
\caption{Slope of the mass function ($\alpha$) for stars within $r_{10}$ (left), stars between $r_{10}$ and $r_m$ (center), and bound stars beyond $r_m$ (right). Models are separated by colour, as indicated in the right panel.}
  \label{fig:alpharad}
\end{figure}

The slope of the mass function in all radial bins (Figure \ref{fig:alpharad}) follows the same trend as the global mass function, however within observational uncertainties the inner mass function  appears to be independent of orbit. The orbital independence is due to two-body interactions being the dominant physical process in the core of a GC relative to tidal stripping. Assuming a Universal IMF, the nearly orbit independent evolution of $\alpha$ for $r < r_{10}$ could be used to solve for the initial MF and hence total initial mass of MW GCs given their core mass function \citep{leigh12}.

For the intermediate mass function, we begin to see a clear separation in the evolution of $\alpha$ for GCs with different orbits. $\alpha$ increases at a slower rate than the inner region, primarily because both two-body relaxation and tidal stripping are in effect. The removal of low mass stars via tidal stripping slows the evolution of $\alpha$ compared to if just two-body relaxation was occurring.

In the outer region we see an initial decrease in $\alpha$ as mass segregation results in high mass stars migrating to the inner region of the GC. However, $\alpha$ quickly begins to increase for tidally filling clusters (e0r6, e05r18) as they lose mass. Unlike the inner region of the cluster, tidal stripping is now the dominant mechanism and can produce significantly different values of $\alpha$ based on cluster orbit. Specifically the difference between e0r6 and e05r18 is larger in the outer region than the intermediate region. With observational uncertainties in $\alpha$ typically ranging from $2$ to $15 \%$ \citep{demarchi10, paust10}, discrepancies of this magnitude should be measurable in high quality observations. For the outer regions of clusters e0r18, e09r104, and e0r104, $\alpha$ is still decreasing as the cluster relaxes. Since outer clusters are either barely tidally filling or not at all (see Figure \ref{fig:rfill}), two-body interaction is the only mechanism affecting the outer region of the GC and the evolution of $\alpha$ is not accelerated due to tidal stripping. Unfortunately, the outer mass functions of Galactic GCs are difficult to measure due to low number statistics and field contamination, and we are forced to rely on mass functions measured near $r_h$. 

In principle, the ratio of $\alpha$ in the core to $\alpha$ in the outskirts could put very tight constraints on orbital eccentricity. Consider two clusters with the exact same mass, $r_h$, $R_{gc}$ and value of $\alpha$ in their outskirts. While one may conclude these two clusters must have similar orbits, this conclusion would be incorrect if the clusters had different sizes or masses at birth. The evolution of $\alpha$ in the core on the other hand is independent of cluster orbit, and only depends on the initial mass and size of the cluster of birth as these properties are what govern the time it takes for the core to relax. Therefore normalizing by the value of $\alpha$ in the core is analagous to normalizing by the initial cluster conditions. In the current example, the cluster with the smaller core $\alpha$ was likely more massive and larger than the other cluster at birth and took longer to relax. To have the same value of $\alpha$ in the outskirts, the cluster with the higher initial mass and size must have an eccentric orbit and be near $R_a$ in order to have lost a higher fraction of its initial mass. Additional simulations of clusters with different initial conditions are required to further explore the usefulness of the ratio of $\alpha$ in the core to $\alpha$ in the outskirts.

\section{Application to Milky Way Globular Clusters} \label{mwapp}

Our models demonstrate that the periodic perigalactic passes and tidal heating experienced by GCs with eccentric orbits can lead to enhanced mass loss, increased velocity dispersions, and shorter relaxation times than if the cluster had a circular orbit at $R_a$, $<R_{gc}>$, $R_\Psi$, or with the same semi-major axis. All of these effects combine to alter the stellar MF of a GC in a predictable manner. Assuming a universal IMF, which is consistent with the results of \citet{leigh12}, the possibility then arises to relate the observationally determined MF of GCs to the tidal field, and thereby constrain GC orbits. A universal IMF is consistent with results of \citet{leigh12}. Below, we use our model results and the MFs of GCs with solved orbits to illustrate how GC orbits can be constrained given $\alpha$ and $R_{gc}$.

\begin{figure}
\centering
\includegraphics[width=\columnwidth]{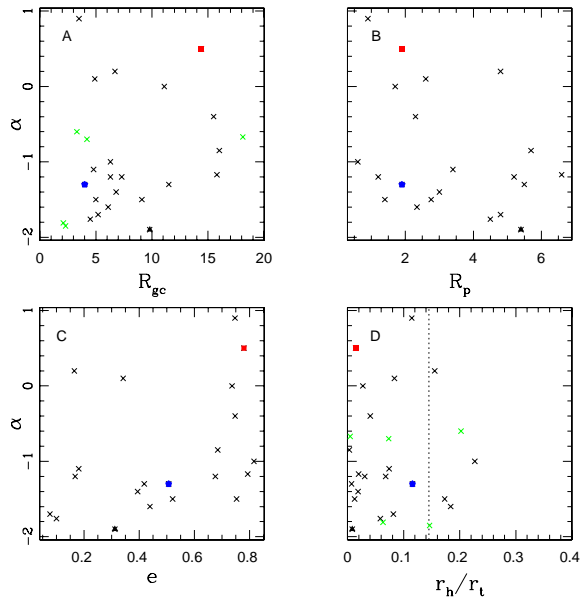}
\caption{Slope of the mass function ($\alpha$) compared to the present $R_{gc}$ (Panel A), $R_p$ (Panel B), orbital eccentricity (Panel C), and $\frac{r_h}{r_t}$ (Panel D)
for Galactic GCs with solved orbits. In Panel D, the vertical line corresponds to $\frac{r_h}{r_t} = 0.145$. NGC 7078 (black triangle), NGC 6809 (blue filled circle) and NGC 2298 (red filled squares) have been highlighted. In Panels A and D, large green crosses mark the clusters in the \citet{demarchi10} dataset with unsolved orbits.}
  \label{fig:mwalpha}
\end{figure}

In Figure \ref{fig:mwalpha}, we plot $\alpha$ from \citet{demarchi10} versus current $R_{gc}$, $R_p$, orbital eccentricity, and the ratio $\frac{r_h}{r_t}$ \citep{harris96} for Galactic GCs with solved orbits \citep{dinescu99, dinescu07, dinescu13}. Cluster tidal radii are calculated at their current $R_{gc}$ given the formalism of \citet{bertin08}. The vertical dotted line in the bottom right panel corresponds to $\frac{r_h}{r_t} = 0.145$, where clusters with $\frac{r_h}{r_t} > 0.145$ are considered to be tidally filling and clusters with  $\frac{r_h}{r_t} < 0.145$ are considered to be tidally under-filling \citep{henon61}. Clusters in the \citet{demarchi10} dataset with unsolved orbits are plotted in Panels A and D as large green crosses. For comparison purposes, NGC 7078 (black triangle), NGC 6809 (blue filled circle) and NGC 2298 (red filled squares) have been singled out as they cover the full range in $R_{gc}$, eccentricity, and $\alpha$. It should be noted that values of $\alpha$ taken from \citet{demarchi10} were measured near the effective radius of the cluster. Therefore differences between eccentric and non-eccentric clusters should follow the behaviour described in the centre panel of Figure \ref{fig:alpharad}.

\subsection{Clusters with Solved Orbits}

\subsubsection{NGC 7078 (M15)}

NGC 7078 (black triangle) has the steepest mass function (most negative $\alpha$) of all the GCs with solved orbits, suggesting it is the least dynamically evolved. Without any prior knowledge about the orbit, this cluster appears at face-value to represent an anomaly as its present day $R_{gc}$ is approximately the mean $R_{gc}$ of all clusters in the dataset. We conclude, based solely on the mass function of NGC 7078, that it has a low orbital eccentricity and a correspondingly large perigalactic distance. Taking into consideration the cluster's known orbital parameters, NGC 7078 actually has one of the largest perigalactic distances of all clusters with solved orbits. Therefore, it experiences a weaker mean tidal field than the majority of GCs. Compared to other clusters with large values of $R_p$, NGC 7078 is very tidally under-filling. Being smaller in size and tidally under-filling, combined with experiencing a weaker mean tidal field, means that NGC 7078 has a very low mass loss rate and has likely retained the majority of its stars. Furthermore, its lower orbital eccentricity means that tidal heating plays a near negligible role.

\subsubsection{NGC 6809 (M55)}

NGC 6809 (blue filled circle) represents the inner most cluster in the dataset with a present day $R_{gc}$ of 4 kpc, however it is less dynamically evolved than one would expect given the strong tidal forces it must experience. Its tidal and effective radii suggest that the cluster has expanded enough such that it is almost tidally filling and stars should be able to be stripped from the outskirts. Therefore we would conclude that the cluster must actually spend more time beyond 4 kpc than within 4 kpc, so it must have a moderate to high orbital eccentricity and be located near $R_p$. This statement is consistent with the solved orbit for this cluster.  The cluster has an orbital eccentricity near 0.5 and an $R_p$ of approximately 2 kpc, meaning that the cluster is currently closer to $R_p$ than $R_a$, such that its current position does not represent the mean tidal field it experiences. The weaker than expected tidal forces experienced by NGC 6809 result in a lower mass loss rate and larger relaxation time, both of which help to account for the relatively unevolved (i.e. steep) slope of the MF. 

\subsubsection{NGC 2298}

Finally, NGC 2298 (red filled squares) is very dynamically evolved as it has an inverted mass function with a large positive value of $\alpha$. Again, without prior orbital information, this cluster would appear to be too dynamically evolved as the weak tidal forces it experiences at its current $R_{gc}$ should not have been able to remove enough stars to invert the mass function. Panel D suggests that NGC 2298 is also very tidally under-filling, so one would expect that it would not be strongly affected by tidal forces. Hence the only way NGC 2298 can be so dynamically evolved given its current $R_{gc}$ would be if it has a highly eccentric orbit that brings it deep into the tidal field of the galaxy. Furthermore, NGC 2298 must be near $R_a$ to explain its extremely low $\frac{r_h}{r_t}$. Our conclusion is confirmed by noting NGC 2298 has an orbital eccentricity of 0.78 (Figure \ref{fig:mwalpha}), a $R_a$ of 15.3 kpc, and a current $R_{gc}$ of 14.4 kpc. Periodic episodes of enhanced mass loss during each perigalactic passes have stripped the majority of low mass stars from the outer regions of NGC 2298 leaving it to appear tidally under-filling when near $R_a$.

Note that a similar argument can be made for NGC 288 and Pal5, which despite having $R_{gc}$'s greater than 10 kpc, both appear to be quite dynamically evolved with an $\alpha$ of 0. With orbital eccentricities greater than 0.68, perigalactic passes bring both clusters deep into the Galactic potential to $R_p$'s less than 2 kpc. Enhanced mass loss and energy injection have accelerated each cluster's evolution compared to if they had circular orbits at their current $R_{gc}$'s.

\subsection{Clusters with Unsolved Orbits}

We have demonstrated that an understanding of how orbital eccentricity can influence the dynamical evolution of GCs can be used to make predictions of a GC's orbit based on its $R_{gc}$ and $\alpha$. While it is difficult to predict cluster orbit based solely on $R_{gc}$ and $\alpha$ without additional simulations to explore possible degeneracies between orbit, initial size, and initial mass, we can make some general statements about the remaining clusters in the \citet{demarchi10} dataset with unsolved orbits (plotted as green crosses in Figure \ref{fig:mwalpha}):

\begin{itemize}

\item NGC 1261 is tidally under- filling, has the largest $R_{gc}$, and has one of the least negative values of $\alpha$ of the remaining clusters suggesting it is similar in nature to NGC 2298. Therefore NGC 1261 is likely located near $R_a$ and has a large (e $>$ 0.7) orbital eccentricity. Its high-e orbit causes NGC 1261 to be subject to significant tidal heating and large injections of energy during perigalactic passes, accelerating its dynamical evolution compared to if it had a circular orbit at its current $R_{gc}$.

\item NGC 6352 and NGC 6496 both have similar values of $\alpha$ to NGC 1261 but are located in the inner region of the MW ( 3 kpc $< R_{gc} < $ 5 kpc). Therefore their orbital eccentricities are likely less than NGC 2298 or NGC 6809 (e $<$ 0.5), and are currently located somewhere between $R_p$ and $R_a$. Since NGC 6352 is tidally filling, it is likely closer to $R_p$. Similarly since NGC 6496 is tidally under-filling it is likely closer to $R_a$.

\item NGC 6304 is tidally filling, but has an extremely negative $\alpha$ considering it is located deep in the galactic potential of the MW ($R_{gc} \sim 2$ kpc). NGC 6304 is comparable to the previously discussed NGC 6809, and likely has a moderate to high (e $\sim$ 0.5) orbital eccentricity and is currently located near $R_p$. Hence its very negative $\alpha$ can be explained by the fact that NGC 6304 spends the majority of its time beyond its current $R_{gc}$.

\item Unfortunately no firm conclusions can be made regarding the orbit of NGC 6541 as it is both extremely tidally under-filling and located at a small $R_{gc}$. Hence the evolution of its mass function is likely independent of its orbit. Its extremely negative $\alpha$ suggests the cluster has retained the majority of its stars over its lifetime and likely formed extremely compact relative to other GCs. Due to its low $R_{gc}$, it is also possible that the cluster is near $R_p$ and has a low eccentricity orbit which brings the cluster slightly farther out in the galactic potential. However the fact that it is so tidally under-filling is surprising given its low $R_{gc}$. It may instead be the case that NGC 6541 is a recently accreted GC or the nucleus of a dwarf galaxy, and did not evolve at its current location in the Milky Way . Further simulations of tidally under-filling clusters on eccentric orbits are required to explore these hypotheses.

\end{itemize}

\section{Summary} \label{summary}

Our simulations show that orbital eccentricity can play an important role in the dynamical evolution of a star cluster. Our models demonstrate that for two GCs located at the same $R_{gc}$, one with a circular orbit and one with an eccentric orbit and $R_a = R_{gc}$, the GC with an eccentric orbit will have:

\begin{itemize}

\item increased mass loss rate
\item smaller size
\item increased velocity dispersion
\item shorter relaxation time
\item shallower mass function

\end {itemize}

The same conclusion would be reached by comparing a cluster with a circular orbit at a smaller $R_{gc}$ to the cluster with a circular orbit at $R_a$. However, the non-static tidal field and periodic perigalactic passes experienced by a cluster with an eccentric orbit produce second order effects.

The first effect of an eccentric orbit is periodic episodes of enhanced mass loss during perigalactic passes. So while the mass loss rate that an eccentric cluster experiences for most of its lifetime may correspond to $<R_{gc}>$, the enhanced episodes of mass loss produce a higher overall mass loss rate. The second effect of perigalactic passes is the energization of inner region stars to larger orbits, as first discussed in \citet{webb13}. The periodic injection of energy into the cluster, combined with additional energy due to tidal heating from a non-static tidal field, increases the kinetic energy of individual stars. Therefore inner region stars will be pushed to larger orbits and stars in the outskirts will be able to escape, decreasing the relaxation time and mass segregation time of the cluster. The combined effects of orbital eccentricity serve to partially balance the decreased tidal field strength the eccentric cluster experiences during the majority of its orbit, such that its evolution is comparable to a cluster with a circular orbit at a distance much less than $R_\Psi$, $<R_{gc}>$, or with the semi-major axis of the eccentric cluster. The recurring example discussed in this paper involves model e09r104, which undergoes a similar dynamical evolution as a cluster with a circular orbit at 18 kpc. 

The influence of tidal heating and perigalactic passes are reflected in the global mass function of eccentric GCs, as it will be flatter (less negative slope) than would be expected given the clusters current $R_{gc}$. A flatter mass function is the direct result of increased tidal stripping of outer region stars that are preferentially low in mass due to mass segregation. Conversely, the inner mass function appears to be independent of cluster orbit as the effects of tidal heating are negligible compared to two-body relaxation. Hence the inner mass functions of Galactic GCs may instead be used to constrain the initial mass and size of the GC, and the ratio of $\alpha$ in the core to $\alpha$ in the outskirts could serve as a tracer of orbital eccentricity.

We make use of the measured mass functions of 33 GCs by \citet{demarchi10}, 28 of which have solved orbits \citep{dinescu99, dinescu07, dinescu13}, to demonstrate how $\alpha$ and $R_{gc}$ can be used to constrain cluster orbit. We then put constraints on the orbital eccentricity of the remaining clusters with unsolved orbits based on their $\alpha$ and $R_{gc}$:

\begin{itemize}

\item NGC 1261 has $e > 0.7$, and is currently located near $R_a$
\item NGC 6352 has $ e < 0.5$, and is currently located near $R_p$
\item NGC 6496 has $ e < 0.5$, and is currently located near $R_a$
\item NGC 6304 has $e \sim 0.5$, and is currently located near $R_p$
\item NGC 6541 is extremely under-filling with a low $R_{gc}$, so its $\alpha$ must be orbit independent. To be under-filling with such a small $R_{gc}$, it is likely that NGC 6541 either formed extremely compact and is currently located near $R_p$ with a low e, is a captured GC, or is a dwarf galaxy remnant.

\end {itemize}

Additional simulations, specifically exploring the influence of orbital inclination, initial size, and initial mass on the dynamical evolution of GCs, will help explain the current dynamical state of all Galactic GCs. Isolating the effects of orbital eccentricity, however, is an important first step towards understanding the different ways tidal heating and periodic perigalactic passes can influence cluster evolution. A complete suite of simulations will allow for specific constraints to be placed on the orbits of GCs that have yet to be solved.

\section*{Acknowledgments}
We would like to thank the referee for constructive comments and suggestions regarding the presentation of the paper. JW, AS, and WEH acknowledge financial support through research grants and scholarships from the Natural Sciences and Engineering Research Council of Canada. JW also acknowledges support from the Dawes Memorial Fellowship for Graduate Studies in Physics.


\bsp

\label{lastpage}

\end{document}